\documentclass[twocol]{ametsoc}

\journal{jas}
\bibpunct{(}{)}{;}{a}{}{,}

\usepackage{textcomp}  
\usepackage{sah_shortcuts}
\usepackage{amsmath,amssymb}

\authors{Spencer A. Hill\correspondingauthor{Spencer Hill, 207B Oceanography, Lamont-Doherty Earth Observatory, 61 Route 9W, Palisades, NY 10964}}
\affiliation{Program in Atmospheric and Oceanic Sciences, Princeton University, Princeton, New Jersey, and Lamont-Doherty Earth Observatory, Columbia University, Palisades, New York}
\email{spencerh@princeton.edu}

\extraauthor{Simona Bordoni}
\extraaffil{Department of Civil, Environmental and Mechanical Engineering (DICAM), University of Trento, Trento, Italy, and Division of Geological and Planetary Sciences, California Institute of Technology, Pasadena, California}

\extraauthor{Jonathan L. Mitchell}
\extraaffil{Department of Atmospheric and Oceanic Sciences, and Department of Earth, Planetary, and Space Sciences, University of California, Los Angeles}

\title{Toward a unified theory for the Hadley cell descending and ascending edges}

\abstract{%
We present theories for the latitudinal extents of both Hadley cells throughout the annual cycle by combining our recent scaling for the ascending edge latitude \citep{hill_solsticial_2021} with the uniform Rossby number (\(\Ro\)), baroclinic instability-based theory for the poleward, descending edge latitudes of \citet{kang_expansion_2012}.  The resulting analytic expressions for all three Hadley cell edges are predictive except for diagnosed values of \(\Ro\) and two proportionality constants.  The theory captures the climatological annual cycle of the ascending and descending edges in an Earth-like simulation in an idealized aquaplanet general circulation model (GCM), provided the descending edge prediction is lagged by one month.  In simulations in this and two other idealized GCMs with varied planetary rotation rate (\(\Omega\)), the winter, descending edge of the solsticial, cross-equatorial Hadley cell scales approximately as \(\Omega^{-1/2}\) and the summer, ascending edge as \(\Omega^{-2/3}\), both in accordance with our theory.
}

\begin{document}
\maketitle

\section{Introduction}
\label{sec:intro}

Climatologically over the annual cycle, the poleward, descending edges of the monthly Hadley cells vary meridionally by \({\lesssim5^\circ}\) latitude about their annual mean positions, considerably less than the \(\sim\)15\degr{}S-15\degr{}N range of the shared, ascending edge \citet[\cf/ Fig.~4 of][]{adam_seasonal_2016}).  Zonally averaged, this results in an intense seasonal cycle of rainfall in the deep tropics versus more consistently dry conditions in the subtropics.  Regional deviations from the zonal average are pronounced---with for example intense Indian summer monsoon rainfall spanning roughly the same latitudes as the Sahara and Arabian deserts \citep{rodwell_monsoons_1996}---nevertheless we focus on the zonal-mean problem, seeking a minimal explanation for these differing annual cycles of the descending and ascending edges (henceforth \(\latd\) and \(\ascentlat\) respectively).

Our starting point for \(\latd\) is the analytical model of \citet[][henceforth K12]{kang_expansion_2012}, whose own starting point is that of \citet[][henceforth H00]{held_general_2000} for the annual-mean \(\latd\) that assumes the Hadley cells terminate where their zonal wind profiles become baroclinically unstable.  KL12 extend the H00 model in two key ways.  First, they generalize from the annual mean to the annual cycle by accounting, albeit diagnostically, for off-equatorial \(\ascentlat\).  For angular-momentum-conserving (AMC) zonal winds as assumed by H00, ascent off the equator results in weaker meridional shears \citep{lindzen_hadley_1988} and thus to baroclinic instability onset occurring farther poleward
than for equatorial ascent.  All else equal, this would cause \(\latd\) to be farther poleward in solsticial seasons when \(\ascentlat\) is farthest from the equator compared to equinoctial seasons.  \citet{hilgenbrink_response_2018} utilize this conceptual framework to understand the response of the Hadley circulation annual cycle to perturbations in ocean heat transports.

Second, KL12 relax the H00 assumption of strictly AMC winds by assuming that the Rossby number (\(\Ro\)) is uniform throughout each Hadley cell but not necessarily unity.  Its formal definition follows below, but for now \(\Ro\) is exactly unity for AMC winds and vanishes for vanishing zonal winds, and KL12 derive an analytical expression for the meridional profile of zonal wind under uniform \({0<\Ro\leq1}\).
  In simulations \citep{walker_eddy_2006} and reanalysis data \citep{schneider_general_2006}, \(\Ro\) is regularly below unity and typically smaller in the equinoctial and summer cells than in the cross-equatorial winter cell \citep{bordoni_monsoons_2008,schneider_eddy-mediated_2008}.  By diagnosing a bulk value of \(\Ro\) for each cell and meteorological season in addition to \(\ascentlat\), KL12 provide closed expressions for the northern and southern hemisphere descending edge latitudes in all four seasons.\footnote{In addition to KL12, other authors have considered a uniform \(\Ro\) in the tropical upper troposphere. \citet{becker_feedback_1997} find that a uniform \({\Ro=0.5}\) approximation (their Fig. 7 and Eq.~28) adequately captures the vorticity distribution in the descending branch of the winter Hadley cell in their simplified, dry GCM.  And  \citet{zurita-gotor_finite-amplitude_2018} discuss the absolute vorticity distribution corresponding to uniform \(\Ro\).}

To advance from diagnosing \(\ascentlat\) toward a closed theory, we recently presented a simple analytical scaling \citep{hill_solsticial_2021} in terms of the thermal Rossby number, which assumes that \(\ascentlat\) is determined by the meridional extent of supercritical radiative forcing into the summer hemisphere.  In essence, in the present study we simply combine that scaling for \(\ascentlat\) with the KL12 model for \(\latd\) (with slightly different assumptions about \(\Ro\) compared to KL12 as detailed below).  The result is a unified theory for \(\ascentlat\) and both hemispheres' \(\latd\) with only two proportionality constants as well as \(\Ro\) diagnosed (with a single \(\Ro\) value for each cell and season).

Separately, in general circulation model (GCM) simulations with differing planetary rotation rates (\(\Omega\)) the solsticial, cross-equatorial Hadley cell expands into both the summer and winter hemispheres as \(\Omega\) decreases.  In eddying atmospheres \(\ascentlat\) is farther poleward than the winter \(\latd\) for slowly rotating cases (\eg/ Fig.~6 of \citet{guendelman_axisymmetric_2018}), whereas in Earth-like cases \(\ascentlat\) and the winter \(\latd\) are nearly the same.\footnote{This does not emerge clearly in axisymmetric simulations, and on theoretical grounds in axisymmetric atmospheres \(\latd\) should be poleward of \(-\ascentlat\) \citep{hill_axisymmetric_2019}.}  We will use our combined theory for \(\ascentlat\) and \(\latd\) along with idealized GCM simulations to help explain this behavior.

In the following sections we:
\begin{itemize}
\item derive and describe fixed-\(\Ro\) zonal wind, angular momentum, and depth-averaged potential temperature fields (Section~\ref{sec:fixed-ro-fields});
\item derive and describe our extension of the KL12 model for \(\latd\) assuming fixed \(\Ro\) and our theory for \(\ascentlat\) (Section~\ref{sec:theory});
\item and test our theory over the annual cycle in a seasonally forced simulation in one idealized GCM and for the solsticial Hadley circulation in simulations at various rotation rates in three idealized GCMs (Section~\ref{sec:sims}).
\end{itemize}
We then conclude with summary and discussion (Section~\ref{sec:conc}).

\section{Uniform-\(\Ro\) fields}
\label{sec:fixed-ro-fields}

In general, absolute angular momentum is
\begin{equation}
M=a\coslat(\Omega a\coslat + u),
\end{equation}
where \(a\) is planetary radius, \(\lat\) is latitude, \(\Omega\) is planetary rotation rate, and \(u\) is zonal velocity.  This can be considered the sum of planetary angular momentum, \({M_\mr{p}(\lat)\equiv\Omega a^2\cos^2\lat}\), and relative angular momentum \(ua\coslat\).  If angular momentum is meridionally uniform and equal to the planetary angular momentum value at the latitude \(\ascentlat\), \(M_\mr{p}(\ascentlat)\), the corresponding AMC zonal wind field is
\begin{equation}
  \label{eq:u-amc}
  \uamc(\lat)=\Omega a\coslat\left(\frac{\cos^2\ascentlat}{\cos^2\lat}-1\right).
\end{equation}
In the context of the Hadley cells, we equate \(\ascentlat\) in \eqref{eq:u-amc} with the cells' ascending edge on the grounds that ascent out of the viscous boundary layer there transmits the local planetary angular momentum, \(M_\mr{p}(\ascentlat)\), to the comparatively inviscid free troposphere \citep{held_nonlinear_1980,lindzen_hadley_1988}.  This neglects the finite width of the ascent branch \citep{watt-meyer_itcz_2019,byrne_dynamics_2019}, but in principle one could compute an effective ascent latitude by averaging the planetary angular momentum over the full extent of the ascending branch, perhaps weighting by the vertical velocity out of the boundary layer at each latitude.

The Rossby number is defined as
\begin{equation}
  \label{eq:ross-num}
  \Ro\equiv-\dfrac{\zeta}{f},
\end{equation}
where \(\zeta\equiv-(a\coslat)^{-1}\pdsl{(u\coslat)}{\lat}\) is relative vorticity and  \(f\equiv2\Omega\sinlat\) is the planetary vorticity (\ie/ the Coriolis parameter).  Absolute vorticity is given by \(\eta=f+\zeta=f(1-\Ro)\).  Absolute vorticity is proportional to the meridional gradient of angular momentum, and as such in an AMC state necessarily \(\eta=0\) and \(\Ro=1\).  In discussing the GCM simulations below, we will make use of a generalized version of \(\Ro\) \citep{singh_limits_2019}, but \eqref{eq:ross-num} is the quantity used in the fixed-\(\Ro\) fields we now define.

From \eqref{eq:ross-num}, \({\zeta=-\Ro f}\).  If \({\Ro<1}\) but horizontally uniform, integrating meridionally yields the fixed-\(\Ro\) zonal wind field,
\begin{equation}
  \label{eq:u-fixed-ro}
  u_\Ro(\lat)=\Ro\uamc(\lat),
\end{equation}
which is simply the AMC zonal wind field scaled by \(\Ro\).  The corresponding angular momentum field is
\begin{equation}
  \label{eq:ang-mom-fixed-ro}
  M_\Ro(\lat)=\Omega a^2\left[(1-\Ro)\cos^2\lat + \Ro \cos^2\ascentlat\right],
\end{equation}
which is a \(\Ro\)-weighted average of the planetary angular momentum at the ascent latitude, \(M_\mr{p}(\ascentlat)\), and the full meridional distribution of the planetary angular momentum, \(M_\mr{p}(\lat)\).

Finally, though it does not enter into our model for the Hadley cell edges below, we present for a Boussinesq atmosphere (see \eg/ Eq.~1 of \citet{hill_axisymmetric_2019} for the full underlying system of equations) the depth-averaged potential temperature field in gradient balance with \(u_\Ro\) (assuming \(u_\Ro\) occurs near the tropopause while \(u\approx0\) at the surface).  Denoted \(\hat\theta_\Ro\), it is given by
\begin{align}
  \label{eq:theta-fixed-ro}
  \dfrac{\hat\theta_\mr{a}-\hat\theta_\Ro(\lat)}{\theta_0}=\dfrac{\Ro}{2\Bu}&\bigg[(2-\Ro)\cos^2\lat +\cos^2\ascentlat\times\\\nonumber
  &\left.\left(4(1-\Ro)\ln\left(\dfrac{\cos\ascentlat}{\cos\lat}\right) + \Ro\dfrac{\cos^2\ascentlat}{\cos^2\lat}-2\right)\right],
\end{align}
where \(\hat\theta_\mr{a}\) is the depth-averaged potential temperature at the latitude \(\ascentlat\), \(\theta_0\) is the Boussinesq reference potential temperature, and
\begin{equation}
  \label{eq:burg-num}
  \Bu\equiv \dfrac{gH}{(\Omega a)^2}
\end{equation}
is the planetary Burger number with gravity \(g\) and tropopause height \(H\).

\begin{figure}[t]
  \centering\noindent
  \includegraphics[width=0.5\textwidth]{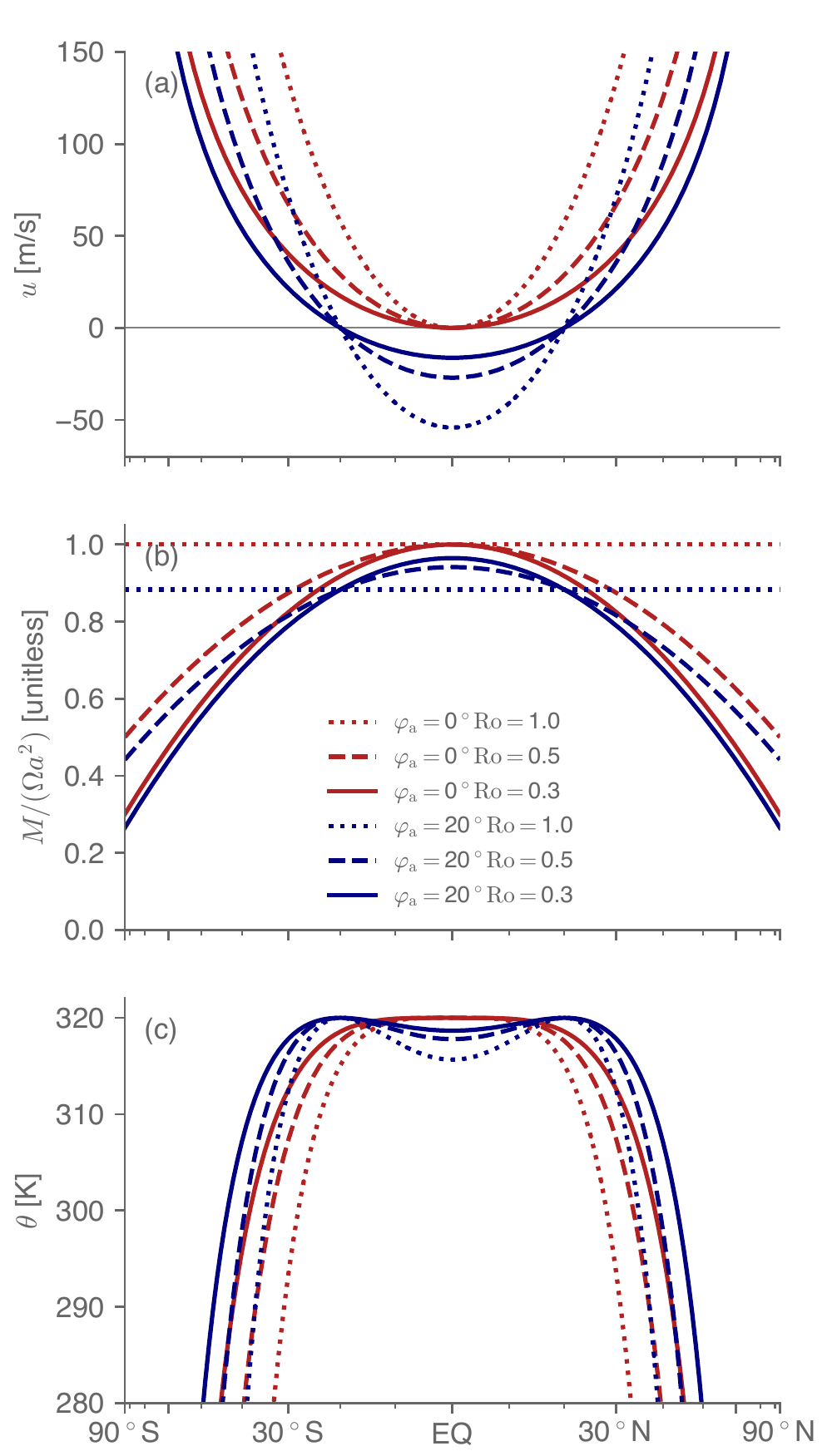}\\
  \caption{Example (a) zonal wind (\(u_\Ro\), in m/s), angular momentum (\(M_\Ro\), normalized by \(\Omega a^2\) and thus unitless), and depth-averaged potential temperature (\(\hat\theta_\Ro\), units Kelvin) fields under a uniform Rossby number.  Red curves are for \(\ascentlat=0\) and blue curves \(\ascentlat=20\)\degr{}S/N, while dotted, dashed, and solid curves are for \({\Ro=1.0}\), 0.5, and 0.3, respectively.  Horizontal axis spacing is in \(\sin\lat\).}
  \label{fig:fixed-ro-examples}
\end{figure}

Fig.~\ref{fig:fixed-ro-examples} shows example \(u_\Ro\), \(M_\Ro\), and \(\hat\theta_\Ro\) fields with \(\Ro=1\), 0.5, or 0.3 and \(\ascentlat=0^\circ\) or 20\degr{}.  For zonal wind, irrespective of \(\Ro\) and \(\ascentlat\), \(u_\Ro\) must vanish at \(\ascentlat\).  If \(\ascentlat=0\), it increases monotonically moving toward either pole, with greater meridional shear the larger \(\Ro\) is.  If \(\ascentlat\neq0\), \(u_\Ro\) is negative from \(-\ascentlat\) to \(\ascentlat\), minimizes at the equator, and increases monotonically toward either pole.  Both moving \(\ascentlat\) off equator and decreasing \(\Ro\) act to weaken the meridional shears.   For example, at 30\degr{}S/N, \(u_\Ro\approx133\), 67, and 40~\ms/ respectively for \(\Ro=1\), 0.5, and 0.3 if \(\ascentlat=0\) or 71, 36, and 21~\ms/ respectively if \({\ascentlat=20^\circ}\).  For angular momentum, irrespective of \(\Ro\) and \(\ascentlat\), \(M_\Ro\) at \(\ascentlat\) is equal to the local planetary angular momentum \(\Omega a^2\cos^2\ascentlat\).  It decreases poleward thereof and, if \(\ascentlat\neq0\), increases equatorward to a maximum at the equator.  For the gradient-balanced potential temperature, \(\hat\theta_\Ro\) maximizes at \(\ascentlat\) irrespective of \(\Ro\) and \(\ascentlat\) and for \({\ascentlat\neq0}\) has a local minimum at the equator.  As \(\Ro\) increases, the meridional temperature gradients increase in magnitude, with a deeper equatorial dip and a more equatorward shoulder poleward of \(\ascentlat\) where temperatures begin dropping rapidly toward the pole.

\section{Combined theory for Hadley cell ascending and descending edges}
\label{sec:theory}

Having defined the fixed-\(\Ro\) fields, we now use them to derive our theory for \(\latd\), which closely follows that of KL12.  We then introduce within it our scaling for \(\ascentlat\), yielding our unified theory for \(\latd\) and \(\ascentlat\).

\subsection{Baroclinic instability onset theory for the Hadley cell edge with \({\Ro<1}\)}
\label{sec:bci-theory}

Following H00, the baroclinic instability criterion for the two-layer model is approximately
\begin{equation}
  \label{eq:bci-crit-2layer}
  \dfrac{u}{\Omega a}=\Bu\deltav\dfrac{\cos\lat}{\sin^2\lat},
\end{equation}
where \(u\) is the zonal wind in the upper layer, the wind in the lower layer has been assumed small enough to neglect, and \(\Deltav\) is a static stability parameter representing the bulk fractional increase in potential temperature from the surface to the tropopause.\footnote{H00 uses the symbol \(R\) to denote the planetary Burger number, which elsewhere \citep{held_nonlinear_1980,hill_axisymmetric_2019} is used for the thermal Rossby number.  To prevent confusion we use the more explicit notation \(\Bu\) for the planetary Burger number and \(\Roth\) for the thermal Rossby number.}\(^,\)\footnote{The tropopause depth \(H\) is assumed horizontally uniform and identical in the RCE state and in the presence of the large-scale circulation; see \citet{hill_axisymmetric_2020} for justification.}  H00 applies this to the annual-mean Hadley cells by assuming on-equatorial ascent, that the zonal winds are AMC, and that the descending edge is identical to this baroclinic instability onset latitude.  Formally, taking \(\ascentlat\approx0\), using (\ref{eq:u-amc}) for \(u\) in \eqref{eq:bci-crit-2layer}, and taking the small-angle limit yields the original H00 theory for \(\latd\), which we denote \(\lat_\mr{H00}\):
\begin{equation}
  \label{eq:h00-edge}
  \lat_\mr{H00}=(\Bu\Deltav)^{1/4}.
\end{equation}
Using \(u_\Ro\) \eqref{eq:u-fixed-ro} rather than \(\uamc\) \eqref{eq:u-amc} as the zonal-wind profile in \eqref{eq:bci-crit-2layer}, for \({\Ro<1}\) the predicted cell edge becomes
\begin{equation}
  \label{eq:h00-fixed-ro}
  \lat_\mr{Ro,ann}=\left(\dfrac{\Bu}{\Ro}\Delta_\mr{v}\right)^{1/4}=\Ro^{-1/4}\lat_\mr{H00},
\end{equation}
with \(\lat_\mr{H00}\) given by \eqref{eq:h00-edge}.  This displaces the cell edge prediction poleward by, for example, \(\sim\)19\% if \({\Ro=0.5}\) or \(\sim\)50\% if \({\Ro=0.2}\).  Using the original \({H=10}\)~km and \({\Deltav=1/8}\) parameter values from \citet{held_nonlinear_1980} yields \({\Bu\approx0.46}\) and \(\lat_\mr{H00}\approx28^\circ\), which becomes approximately 33.3\degr{} if \({\Ro=0.5}\) or 41.9\degr{} if \({\Ro=0.2}\).  This poleward displacement as \(\Ro\) decreases adheres to physical intuition: because the upper-layer zonal wind magnitude at each latitudes decreases as \(\Ro\) decreases, the two-layer baroclinic instability onset criterion is met farther poleward.

For \(\ascentlat\neq0\), using \eqref{eq:u-fixed-ro} in \eqref{eq:bci-crit-2layer} gives without approximation
\begin{equation}
  \label{eq:edge-solst-nosmall}
  \dfrac{\sin^4\latd}{\cos^2\latd}-\sin^2\ascentlat\dfrac{\sin^2\latd}{\cos^2\latd}-\dfrac{\mr{Bu}\Deltav}{\Ro}=0,
\end{equation}
where \(\latd\) is the descending edge.  From \eqref{eq:h00-fixed-ro} the last term could equivalently be written \(-\lat_\mr{Ro,ann}^4\).  These arguments serve equally for the cross-equatorial, winter cell and the summer cell (provided it exists), depending on which cell the chosen \(\Ro\) value is representative of.  A corollary is that if the mean \(\Ro\) value is the same in both cells, then the Hadley circulation extends equally far into either hemisphere irrespective of \(\ascentlat\).  Note that our assumption of uniform \(\Ro\) throughout either cell differs slightly from KL12, who assume \({\Ro=1}\) from the summer-hemisphere edge of the cross-equatorial cell to the equator, a uniform \(\Ro\) value in the winter hemisphere, and a uniform \(\Ro\) value throughout the summer cell.

\begin{figure}[t]
  \centering\noindent
  \includegraphics[width=0.5\textwidth]{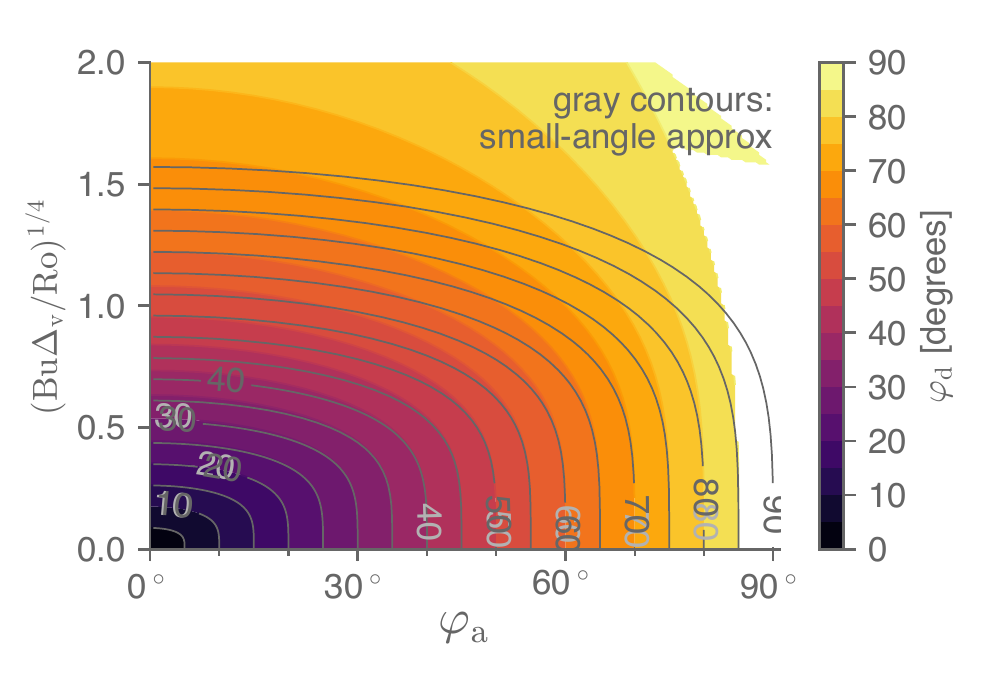}\\
  \caption{Numerical solutions of \eqref{eq:edge-solst-nosmall} for values of \(\ascentlat\) and of \((\Bu\deltav/\Ro)^{1/4}\), with \(\ascentlat\) sampled from 0 to 90 degrees latitude in 1-degree increments and \((\Bu\deltav/\Ro)^{1/4}\) (which is dimensionless) from 0 to 2.0 in 0.01 increments.  Areas in white indicate that the simple numerical algorithm used did not converge.  Contours are from 5 to 90\degr{} in 5\degr{} increments according to the colorbar and also labeled in light gray.  Overlaid dark gray contours and labels are the corresponding small-angle solutions obtained using \eqref{eq:edge-solst}.}
  \label{fig:latd-2d-sweep}
\end{figure}

We have performed a 2D parameter sweep over \(\ascentlat\) and \((\Bu\deltav/\Ro)^{1/4}\), from 0\degr{} to 90\degr{} in 1\degr{} increments for \(\ascentlat\) and from 0 to 2 in 0.01 increments for \((\Bu\deltav/\Ro)^{1/4}\), solving \eqref{eq:edge-solst-nosmall} numerically for each pair of parameter values.  The results are shown as shaded contours in Fig.~\ref{fig:latd-2d-sweep}.  The value of \(\latd\) increases monotonically with \(\ascentlat\) and with \(\lat_\mr{Ro,ann}\).    Close to the vertical axis of Fig.~\ref{fig:latd-2d-sweep}, \({|\lat_\mr{Ro,ann}/\ascentlat|\gg1}\), and \({\latd\approx\lat_\mr{Ro,ann}}\): \(\ascentlat\) is negligibly off-equator.  Close to the horizontal axis of Fig.~\ref{fig:latd-2d-sweep}, \({|\lat_\mr{Ro,ann}/\ascentlat|\ll1}\), and thus \(\latd\approx\ascentlat\).  This regime usefully describes cases where the summer cell effectively disappears, as in all of the perpetual solstice simulations we will discuss below.  For the winter hemisphere, the interpretation is that baroclinic instability onset occurs just poleward of \(-\ascentlat\) near enough that it can be approximated as \(-\ascentlat\).  For intermediate values, if \eg/ \({\ascentlat\approx\lat_\mr{Ro,ann}}\), then \(\latd\) is displaced 27\% poleward of \(\ascentlat\).  We note that \(|\latd|\geq|\ascentlat|\); the descending edge latitude is always at or poleward of the ascending edge latitude.  This is as it should be for the summer-hemisphere \(\latd\) but will prove imperfect for the winter \(\latd\) in the idealized GCM simulations discussed below.

Making the small-angle approximation in \eqref{eq:edge-solst-nosmall} for \(\latd\) and \(\ascentlat\) results in closed expression for \(\latd\):
\begin{equation}
  \label{eq:edge-solst}
  \latd^2=\ascentlat^2\left(\dfrac{1}{2}+\sqrt{\dfrac{1}{4}+\dfrac{\Bu\deltav}{\Ro\ascentlat^4}}\right)=\ascentlat^2\left(\dfrac{1}{2}+\sqrt{\dfrac{1}{4}+\left(\dfrac{\lat_\mathrm{Ro,ann}}{\ascentlat}\right)^4}\right),
\end{equation}
This small-angle approximation is shown as overlaid contours in Fig.~\ref{fig:latd-2d-sweep}.  Only for \((\Bu\Deltav/\Ro)^{1/4}\gtrsim0.8\) does the small-angle approximation error exceed a few degrees latitude, irrespective of \(\ascentlat\).


\subsection{Incorporating theory for \(\ascentlat\)}
\label{sec:ascent-theory}

Using (\ref{eq:edge-solst}) requires knowledge of \(\ascentlat\), which KL12 diagnose.  \citet{hill_solsticial_2021} derive a prognostic theory for \(\ascentlat\) based on the meridional extent of supercritical forcing, \ie/ the latitude range within the summer hemisphere for which the gradient-balanced wind fields that would emerge absent a large-scale circulation given the forcing would be symmetrically unstable.  Specifically,
\begin{equation}
  \label{eq:ascent-hides}
  \ascentlat=c_\mr{a}\left(\dfrac{\Roth}{2}\right)^{1/3},
\end{equation}
where \(c_\mr{a}\) is an empirically determined fitting parameter and
\begin{equation}
  \label{eq:therm-ross-num}
  \Roth\equiv\Bu\deltah\sin\maxlat
\end{equation}
is the thermal Rossby number, with \(\deltah\) a parameter of the imposed forcing field that determines the bulk meridional temperature gradients of the forcing in conjunction with \(\maxlat\), the latitude at which the forcing maximizes.   For the solsticial, cross-equatorial Hadley cells in the simulations analyzed by \citet{hill_solsticial_2021} the best-fit value of \(c_\mr{a}\) ranges from 1.3 to 2.6 across three idealized GCMs.\footnote{\citet{hill_solsticial_2021} report values for \(c_\mr{a}\) of 1.0, 1.7, and 2.1 for the \citet{faulk_effects_2017}, \citet{singh_limits_2019}, and \citep{hill_solsticial_2021} simulations, respectively, but these implicitly incorporate the \({2^{-1/3}\approx0.8}\) factor in \eqref{eq:ascent-hides}.  We separate it out from \(c_\mr{a}\) for better consistency with  \eqref{eq:ascent-hides}.}  For \(\Roth\), \citet{hill_solsticial_2021} show that for solsticial seasons one can attain an accurate estimate with \(\maxlat\) set to 90\degr{} by tuning the value of \(\deltah\).  Doing so, the non-standard \(\sin\maxlat\) term drops out and the \(\Roth\) definition becomes the more conventional \(\Roth=\Bu\deltah\).  But the \(\sin\maxlat\) dependence is necessary for understanding the annual cycle as will be discussed further below.

Given diagnosed values of \(c_\mr{a}\) and \(\Ro\) for each Hadley cell, \eqref{eq:ascent-hides} in conjunction with \eqref{eq:edge-solst} provide a theory for all three cell edges: \eqref{eq:ascent-hides} predicts \(\ascentlat\), and using that in \eqref{eq:edge-solst} for \(\latd\) then yields
\begin{equation}
  \label{eq:edge-solst-crit}
  \latd=c_\mr{d}c_\mr{a}\left(\dfrac{\Roth}{2}\right)^{1/3}\sqrt{\dfrac{1}{2}+\sqrt{\dfrac{1}{4}+\dfrac{2^{4/3}}{c_\mr{a}^4}\dfrac{\deltav}{\deltah\sin\maxlat}\dfrac{1}{\Ro\Roth^{1/3}}}},
\end{equation}
where we have also included the empirical fitting parameter \(c_\mr{d}\) that will prove necessary for the simulations analyzed below.  The term \(\Deltav/\Deltah\sin\maxlat\) amounts to a seasonally varying bulk isentropic slope of the radiative-convective equilibrium state.  From \eqref{eq:edge-solst-crit}, the poleward edge of either Hadley cell increases with increasing \(\Roth\), increasing isentropic slope, or decreasing \(\Ro\).  Large \(\Roth\) corresponds to the large-\(\ascentlat\) limit above, \({\latd\approx\ascentlat}\), while small \(\Roth\) corresponds to the small-\(\ascentlat\) limit of \({\latd\approx\lat_\mr{Ro,ann}}\).


\section{Simulation results}
\label{sec:sims}

We now assess the above theoretical arguments against simulations in two moist GCMs and one dry idealized GCM.  After describing the models and simulations, we consider the annual cycles of \(\latd\) and \(\ascentlat\) in an Earth-like aquaplanet control simulation, followed by their behaviors across a wide range of rotation rates in all three GCMs.

\subsection{Description of models and simulations}
Details of the model formulations and simulations are provided by \citet{hill_solsticial_2021}.  Briefly, the dry model \citep{schneider_tropopause_2004} approximates radiative transfer via Newtonian cooling, with the equilibrium temperature field that temperatures are relaxed toward being the forcing field from \citet{lindzen_hadley_1988} but maximizing at the north pole.  The relaxation field is statically unstable, and a simple convective adjustment scheme relaxes temperatures of unstable columns toward a specified lapse rate at a fixed timescale of \(\gamma\Gamma_\mr{d}\), where \(\Gamma_\mr{d}=g/\cp\) is the dry adiabatic lapse rate with \(c_p\) the specific heat of dry air at constant pressure, and \(\gamma=0.7\) mimics the stabilizing effects of latent heat release that would occur in a moist atmosphere (though the model is otherwise dry).  Four simulations are performed, three with the \(\deltah\) parameter that determines the horizontal temperature gradients of the forcing set to 1/15 and with the planetary rotation rate set to 0.25, 1, or 2\(\times\) Earth's value, and another with \(\deltah=1/6\) and Earth's rotation rate.

The moist simulations are those originally presented by \citet{faulk_effects_2017} and \citet{singh_limits_2019}.  The \citet{faulk_effects_2017} simulations use the idealized aquaplanet model of \citet{frierson_gray-radiation_2006} with a slab-ocean lower boundary with a 10-m mixed layer depth.  They are forced either with an annual cycle of insolation approximating that of present-day Earth, or with insolation fixed at northern solstice.  The annual cycle simulations include planetary rotation rates ranging from 1/32 to 4\(\times\) Earth's by factors of two, while the three perpetual solstice simulations are at 1, 1/8, or 1/32\(\times\) Earth's rotation rate.  The \citet{singh_limits_2019} simulations use an idealized aquaplanet close to that of \citet{ogorman_hydrological_2008}, itself a modified version of the \citet{frierson_gray-radiation_2006} model.  All of these simulations use a time-invariant, solsticial insolation forcing as in the second subset of the \citet{faulk_effects_2017} simulations, with rotation rates ranging from 1/8 to 8\(\times\) Earth's.

The simulated values of \(\latd\) are diagnosed conventionally as the latitude at which the mass-overturning streamfunction at the level of the cell center reaches 10\% of its maximum value, with an additional \(\cos\lat\) weighting factor that accounts for constricting latitude circles moving poleward \citep{singh_limits_2019}.  The 10\% threshold is needed rather than a zero crossing for cases with large Hadley cells, in which the Ferrel cells and/or summer Hadley cell can be nonexistent and the streamfunction same-signed (albeit very weak) all the way to either pole.  For \(\ascentlat\), the same 10\% threshold is used in the perpetual-solstice simulations and in the annual cycle simulations for months in which the summer Hadley cell has effectively vanished.  In months where both Hadley cells are well defined, \(\ascentlat\) is taken as the average of the inner edges of the two cells computed using this 10\% criterion (which is approximately the latitude of the streamfunction zero crossing; not shown).

\subsection{Annual cycles of \(\ascentlat\) and \(\latd\)}

KL12 apply their theory to the seasonal (DJF, MAM, JJA, and SON) Hadley cell descending edges in CMIP3 GCMs, arguing that \(\Ro\) and \(\ascentlat\) compensate with respect to \(\latd\): at solstice, \(\ascentlat\) is farthest poleward, which acts to move both the summer and winter \(\latd\) poleward, but \(\Ro\) is closest to unity in the solsticial cross-equatorial cell, which acts to move the winter \(\latd\) equatorward.  In our formalism, by \eqref{eq:edge-solst-crit} if \(\ascentlat\) is small relative to \({\lat_\mr{Ro,ann}=(\Bu\deltav/\Ro)^{1/4}}\) throughout the annual cycle, then the annual cycle of \(\latd\) is determined by the annual cycle of \(\Ro\) (provided \(H\) that appears in \(\Bu\) and \(\deltav\) are constant across seasons).  We now asses the seasonally forced simulation at Earth's rotation rate of \citet{faulk_effects_2017}, which turns out to follow effectively the opposite limit: \(\ascentlat\) variations (which are well predicted by supercriticality) with \({\Ro=1}\) assumed throughout the annual cycle account for the annual cycle of the winter \(\latd\).

\begin{figure}[t]
  \centering\noindent
  \includegraphics[width=0.5\textwidth]{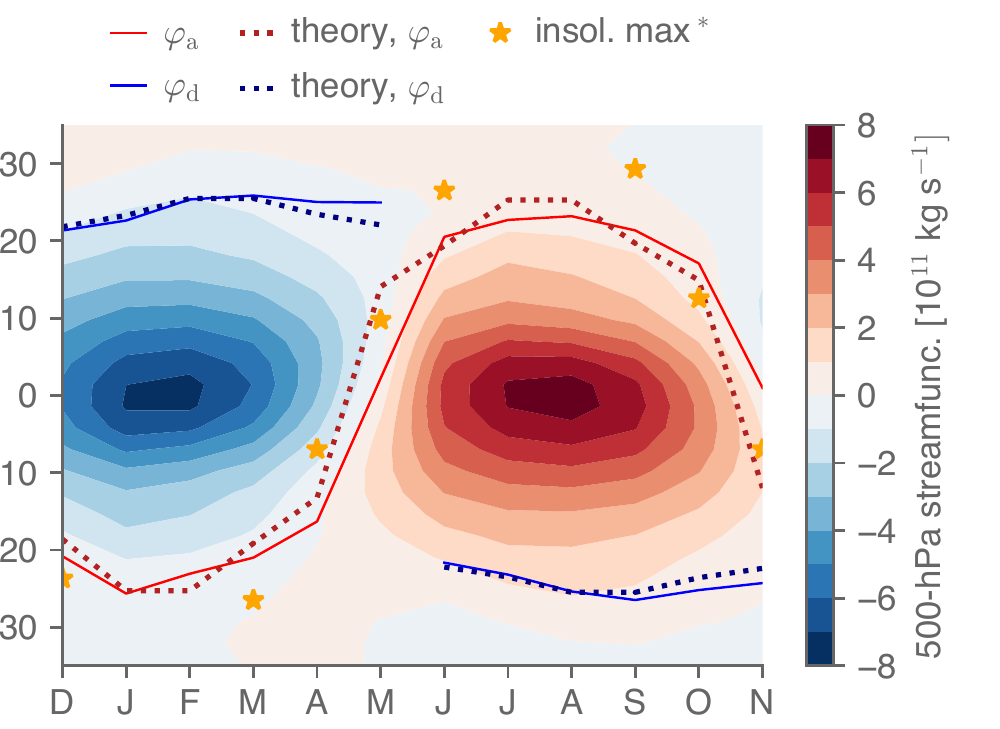}\\
  \caption{In the seasonally forced, Earth-like aquaplanet simulation, climatological annual cycle of Hadley cell streamfunction at 500~hPa in shading according to the colorbar, as well as Hadley cell edges and theories for the Hadley cell edges as indicated in the legend.  Note that the insolation maximum (orange stars) is shifted by one month to facilitate comparison with \(\ascentlat\), and in the two months nearest solstice the maximum is near the summer pole and thus not shown.}
  \label{fig:ann-cyc-f17}
\end{figure}

Fig.~\ref{fig:ann-cyc-f17} shows the climatological annual cycles of \(\ascentlat\), \(\latd\) in both hemispheres, and the meridional overturning streamfunction at 500~hPa, as well as theoretical estimates described below for each cell edge.  The simulated cells are reasonably Earth-like in their meridional extent, overturning strength, and annual cycle phasing.  In particular, \(\ascentlat\) (solid red curve) migrates into either summer hemisphere but with a roughly 1-month lag behind the insolation, and its seasonal migrations (25.7\degr{}S-23.2\degr{}N) are several times larger than those of \(\latd\), (solid blue curves; 21.3-30.0\degr{}N and 21.7-26.5\degr{}S).  Because the summer-hemisphere cell is so weak, \(\latd\) is only well defined in the winter half-month for either hemisphere (December through May for the northern hemisphere, June through November for the southern hemisphere).  Compared to Earth, the \(\ascentlat\) excursions into the summer hemisphere and cross-equatorial Hadley cell overturning strength are moderately excessive, while the simulated summer cell is too weak over most of the annual cycle.  Finally, the transition from equinoctial to solsticial regimes is excessively rapid, approaching closer to the square-wave prediction of \citet{lindzen_hadley_1988} for axisymmetric atmospheres than Earth's more sinusoidal variations \citep{dima_seasonality_2003}; in fact the annual-mean (not shown) rainfall and Hadley cells both show a double ITCZ resulting from this rapid jumping of the ascent rather deep into either summer hemisphere.  These discrepancies are likely due to the rather shallow 10-m mixed layer, leading to excessively rapid seasonal transitions \citep{donohoe_effect_2014,wei_energetic_2018}.

The insolation maximum (orange stars, lagged by 1 month for ease of comparison with \(\ascentlat\)), which corresponds to the forcing maximum latitude \(\maxlat\), occurs near the summer pole during the core solsticial months, whereas \(\ascentlat\) (solid red curve) stays within the tropics.  In the other months of the year, \(\ascentlat\approx\maxlat\).  The supercriticality scaling for \(\ascentlat\) \eqref{eq:ascent-hides} captures this.  Specifically, we set \({\deltah=1/15}\) uniformly throughout the annual cycle and take \(\maxlat\) to be the insolation maximum latitude, lagged by 1 month to accommodate the system's thermal inertia.\footnote{\citet{hill_axisymmetric_2020} diagnose \(\deltah\) from latitude-by-latitude RCE simulations under annual-mean insolation and find \(\deltah\approx1/8\), roughly twice that of the \(\deltah\approx1/15\) value for RCE simulations under solsticial forcing from \citep{hill_solsticial_2021}.  Presumably then \(\deltah\) would be even larger under equinoctial forcing.  But we do not attempt to account for this seasonality in \(\deltah\).}  Setting \({c_\mr{a}\approx1.8}\) for all months then yields a good approximation (dotted dark red curve) to the simulated \(\ascentlat\).  This value of \(c_\mr{a}\) is \(\sim\)37\% larger than the best-fit value of 1.31 for the solsticial \(\ascentlat\) across the \citet{faulk_effects_2017} seasonally forced simulations with different rotation rates---a neither trivial nor order-of-magnitude difference, suggesting that the proportionality is moderately influenced by different processes in these two distinct contexts.  The 1.8 value is also less than the values of 2.2 and 2.6 diagnosed across rotation rates for, respectively, the simulations of \citet{singh_limits_2019} and the dry simulations of \citet{hill_solsticial_2021}.

We then use \(\ascentlat\) predicted as just described to predict \(\latd\) as follows.  Due to an inadvertent loss of zonal-wind data from the \citet{faulk_effects_2017} simulations, we are not able to directly diagnose \(\Ro\).  Instead, we assume \(\Ro=1\), which provided \(0\leq\Ro\leq1\) yields the equatorward-most possible \(\latd\) predictions, all else equal.  Even still, this yields a \(\latd\) prediction poleward of the simulated \(\latd\) values (not shown), which we correct for by setting \({c_\mr{d}=0.75}\) in \eqref{eq:edge-solst-crit}.  We then shift the results later in time by 1-month, resulting in a considerably accurate fit to the simulations (dotted blue curve).  In the concluding section below we provide speculative arguments to justify this equatorward displacement and 1-month phase lag of \(\latd\) compared to \(\ascentlat\).  The \(\latd\) is only marginally improved if the actual simulated \(\ascentlat\) values are used rather than our predicted \(\ascentlat\) (not shown).

These results suggest that relatively muted annual cycles of \(\latd\) relative to that of \(\ascentlat\) can emerge via different mechanisms even restricting to a reasonably Earth-appropriate segment of the parameter space.  In the comprehensive GCMs analyzed by KL12, \(\ascentlat\) presumably varies closer to the real-world value and thus less than in our aquaplanet simulations, while the seasonal \(\Ro\) values that KL12 indirectly diagnose as a fitting parameter range over 0.45--1.  By \eqref{eq:edge-solst}, using \({H=10}\)~km, then for Earth \({\Bu\approx0.46}\), which with the standard \({\deltav=1/8}\) value, \(\ascentlat\) ranging from the equator to 15\degr{}S/N, and these \(\Ro\) ranges yields a maximal \(\latd\) range of 28.0\degr{} (\({\ascentlat=0}\), \({\Ro=1}\)) to 35.9\degr{} (\({\ascentlat=15^\circ}\), \({\Ro=0.45}\)).  But these values are fairly insensitive to both \(\Ro\) and \(\ascentlat\) within these ranges: if \(\Ro\) is kept at unity, the \({\ascentlat=15^\circ}\) prediction moves equatorward only by 1.7\degr{}, and conversely if \({\ascentlat}\) is at the equator the \({\Ro=0.45}\) prediction moves poleward by only 2.1\degr{}.  Conversely, in our simulation the monthly variations of \(\latd\) are comparable to the CMIP3 GCMs, \({\lesssim5^\circ}\) about their annual means, but---with \(c_\mr{d}\) and the 1-month lag from \(\ascentlat\) accounted for---appear determined almost entirely by the seasonality of \(\ascentlat\) with \(\Ro\) constant.

\subsection{Relative behaviors of solsticial \(\ascentlat\) and \(\latd\) across rotation rates}

In the small-\(\Roth\) limit, \eqref{eq:ascent-hides} predicts \(\ascentlat\propto\Roth^{1/3}\), and \citet{hill_solsticial_2021} show that this accurately describes the solsticial \(\ascentlat\) across planetary rotation rates in the idealized GCM simulations presently under consideration.  For \(\latd\), by \eqref{eq:edge-solst-nosmall} for small \(\Roth\) and thus small \(\ascentlat\),
\begin{equation}
  \label{eq:latd-scaling}
  \latd\approx c_\mr{d}\left(\dfrac{\Bu\deltav}{\Ro}\right)^{1/4}=c_\mr{d}\left(\dfrac{\Roth}{\Ro}\dfrac{\deltav}{\deltah}\right)^{1/4},
\end{equation}
again incorporating the empirical fitting parameter \(c_\mr{d}\).  In that case, provided \(\Ro\) does not change appreciably then \({\latd\propto\Bu^{1/4}\sim\Roth^{1/4}}\), where in this context we can substitute \(\Roth\) for \(\Bu\) since only \(\Omega\) is varied and appears identically (as \(\Omega^{-2}\)) in the two nondimensional numbers.  We now argue that the idealized GCM simulations reflect this modest \({1/3-1/4=1/12}\) difference in power-law exponent in \(\Roth\) for \(\ascentlat\) \vs/ \(\latd\).

\begin{figure}[t]
  \centering\noindent
  \includegraphics[width=0.5\textwidth]{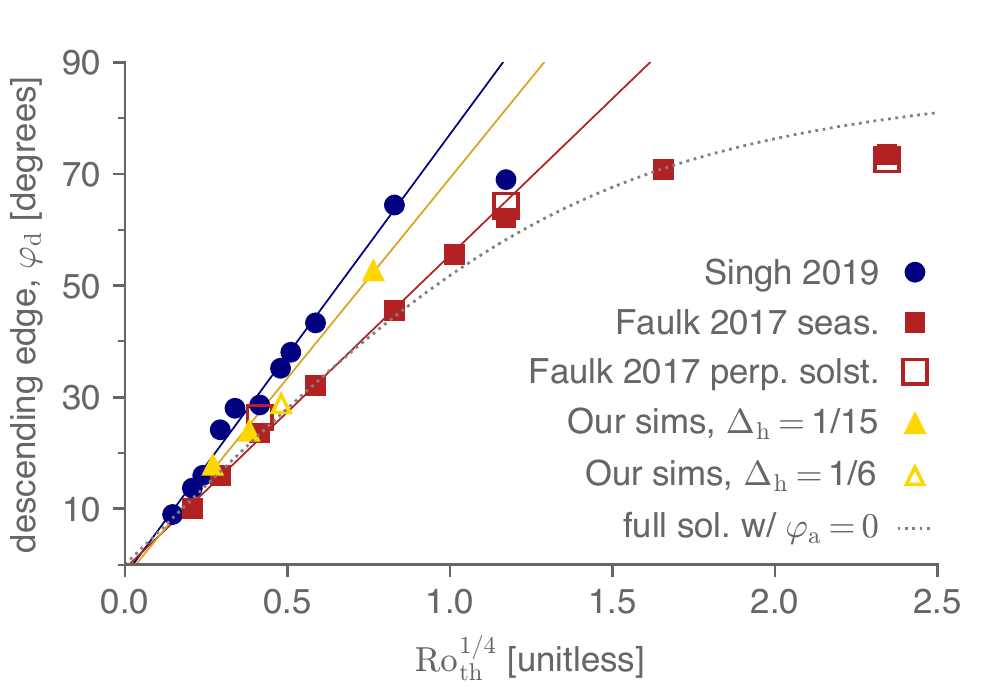}\\
  \caption{Latitude of the winter-hemisphere descending edge of the cross-equatorial Hadley cell, \(\latd\), in idealized aquaplanet simulations of \citet{faulk_effects_2017}, \citet{singh_limits_2019}, and in the idealized dry simulations of \citet{hill_solsticial_2021} as a function of the thermal Rossby number to the one-fourth power, each signified by different symbols as indicated in the legend.  The solid lines show the linear best fit for \(\latd\) as a function of \(\Roth^{1/4}\) for the given simulation set, restricting to \({\Roth<1}\), with red, blue, and yellow for the \citet{singh_limits_2019}, \citet{faulk_effects_2017}, and the \({\deltah=1/15}\) dry simulations respectively.  The dotted gray curve is the numerical solution to \eqref{eq:edge-solst-nosmall} with \({\ascentlat=0}\) and \({\Ro\deltah/\deltav=1}\).}
  \label{fig:sims}
\end{figure}

Fig.~\ref{fig:sims} shows the winter \(\latd\) for all the simulations as a function of \(\Roth^{1/4}\).  For each model, a best-fit line is included of \(\latd\) with \(\Roth^{1/4}\) for simulations with \(\Roth<2\).  Overall the simulations follow this scaling well.  Table~\ref{table:power-fits} lists the slope and intercept from the linear best fits of \(\latd\) against \(\Roth^{1/4}\), with the slope amounting to a best fit for the empirical \(c_\mr{d}\) parameter in \eqref{eq:edge-solst-crit} (at least in the small-\(\Roth\) limit).  The inferred \(c_\mr{d}\) values range from 1.4 for the \citet{singh_limits_2019} simulations to 0.9 for the \citet{faulk_effects_2017} perpetual solstice simulations.  The value of 1.0 for the \citet{faulk_effects_2017} seasonal cycle simulations is modestly higher than the value discussed above of 0.8 for the climatological annual cycle in the \citet{faulk_effects_2017} simulation at Earth's rotation rate---opposite to \(c_\mr{a}\), which was larger for the annual cycle than across rotation rates.  The intercepts, which in theory should be zero, range from -2.4 to 5.6\degr{} latitude and average across the simulation sets to a modest 0.3\degr{}.

Table~\ref{table:power-fits} includes best-fit power law exponents for \(\latd\) and for \(\ascentlat\) against \(\Roth\) computed for each set of simulations by linear regression in log-log space.  For all sets of simulations, the inferred exponent is larger and closer to 1/3 for \(\latd\) than for \(\ascentlat\), which is closer to 1/4.  The dry simulations exhibit the largest exponents for both, 0.41 and 0.3, respectively, and the \citet{faulk_effects_2017} seasonally forced and perpetual-solstice simulations, respectively, give the smallest exponents, 0.28 and 0.21.  The average of the best-fit exponents across the four simulation sets are nearly identical to the scalings, 0.26 and 0.33.

\begin{table*}
  \begin{center}
    \begin{tabular}{lrrr|rrr}
      \hline\hline
      &\multicolumn{3}{c}{winter \(\latd\)} & \multicolumn{3}{c}{summer \(\ascentlat\)}\\
      & power & \(c_\mr{d}\) & intercept &  power & \(c_\mr{a}/2^{1/3}\) & intercept \\
      \hline
      theory            & 0.25 & n/a &    0\degr{} & 0.33 & n/a  &    0\degr{} \\
      S19               & 0.25 & 1.4 & -1.9\degr{} & 0.32 & 1.8 & -2.3\degr{}\\
      F17, seas. forc.  & 0.26 & 1.0 &  5.6\degr{} & 0.28 & 1.0 &  4.2\degr{}\\
      F17, perp. solst. & 0.21 & 0.9 & -0.7\degr{} & 0.30 & 1.1 &  4.4\degr{}\\
      dry, LH88-forced  & 0.30 & 1.3 & -2.4\degr{} & 0.41 & 2.1 & -9.2\degr{}\\
      F17, ann. cyc.    & n/a  & 0.8 & n/a         & n/a  & 1.8 & n/a \\
      \hline
    \end{tabular}
    \caption{Best-fit exponents of power law scalings for the winter and summer edges of the cross-equatorial solsticial Hadley cell in each set of simulations, as well as the best-fit slope and intercepts for each simulation set against the theoretical \(\Roth\) power law.  The slope for \(\latd\) amounts to an approximation of \(c_\mr{d}\) and that of \(\ascentlat\) an approximation of \(c_\mr{a}\); the latter is reported with the additional \(2^{-1/3}\) factor included to facilitate direct comparison with \(c_\mr{d}\).  Simulations are restricted to those for which \({\Roth<2}\), since the theoretical predictions of 1/3 and 1/4 for the winter and summer edges, respectively, assume small angle and thus small \(\Roth\).  The dry LH88-forced simulations do not include the \(\deltah=1/6\) case.  S19 stands for \citet{singh_limits_2019}, and F17 stands for \citet{faulk_effects_2017}.  The last row lists the diagnosed \(c_\mr{d}\) and \(c_\mr{a}\) values for the annual cycle in the \citet{faulk_effects_2017} Earth-like simulation.}
    \label{table:power-fits}
  \end{center}
\end{table*}

As \(\Roth\) increases beyond \(\sim\)1, the simulated \(\latd\) level off, never exceeding \(\sim\)70\degr{}.  The full, non-small-angle expression \eqref{eq:edge-solst-nosmall} solved numerically with \(\ascentlat=0\) and all parameters except \(\Omega\) set to Earth-like values (dotted grey curve) qualitatively captures the leveling off of \(\latd\).  This \(\sim\)70\degr{} contrasts with \(\ascentlat\), which in the slowly-rotating regime is near the summer pole \citep{hill_solsticial_2021}.  Table~\ref{table:power-fits} includes the slope and intercepts of \(\ascentlat\) against \(\Roth^{1/3}\), with the slope amounting to a best fit for \(c_\mr{a}/2^{1/3}\) in \eqref{eq:ascent-hides}.  The ratio \(2^{1/3}c_\mr{d}/c_\mr{a}\) takes values approximately over 0.6-1 in the four simulation sets.  If \(\Ro\) and the forcing isentropic slope \(\deltav/\deltah\) are both nearly unity, it can be shown from \eqref{eq:ascent-hides} and \eqref{eq:latd-scaling} that \(\ascentlat=\latd\) if \(2^{1/3}c_\mr{d}/c_\mr{a}=\Roth^{1/12}\).  Given the 0.6-1 range for the left-hand-side, this yields \(\Roth\approx1\), with \(\latd\) equatorward of \(\ascentlat\) for larger \(\Roth\).  This roughly captures the behavior of the simulations.

The unfilled yellow triangle in Fig.~\ref{fig:sims} shows the dry, LH88-forced simulation at Earth's rotation rate and with \(\deltah=1/6\) rather than \(\deltah=1/15\) as in the other three dry simulations.  As is the case for \(\ascentlat\) \citep{hill_axisymmetric_2020}, \(\latd\) is somewhat separated from the power law of the \(\deltah=1/15\) cases.  Strictly speaking, in the \(\ascentlat\approx0\) limit of \eqref{eq:edge-solst-nosmall}, \(\latd\) is independent of \(\deltah\).  But, while small, \(\ascentlat\neq0\) in the simulations, and since an increase in \(\deltah\) moves \(\ascentlat\) poleward, it is qualitatively consistent that \(\latd\) moves poleward as a result.  Given that the annual cycle amounts to a variation in \(\deltah\sin\maxlat\), it is worth noting that the slope between the \(\deltah=1/15\) and \(\deltah=1/6\) cases at Earth's rotation rate is shallower than that inferred across rotation rates at \(\deltah=1/15\), which qualitatively coheres with \(c_\mr{d}\) being smaller for the annual cycle than across rotation rates in the \citet{faulk_effects_2017} simulations (Table~\ref{table:power-fits}).

\begin{figure*}[t]
  \centering\noindent
  \includegraphics[width=\textwidth]{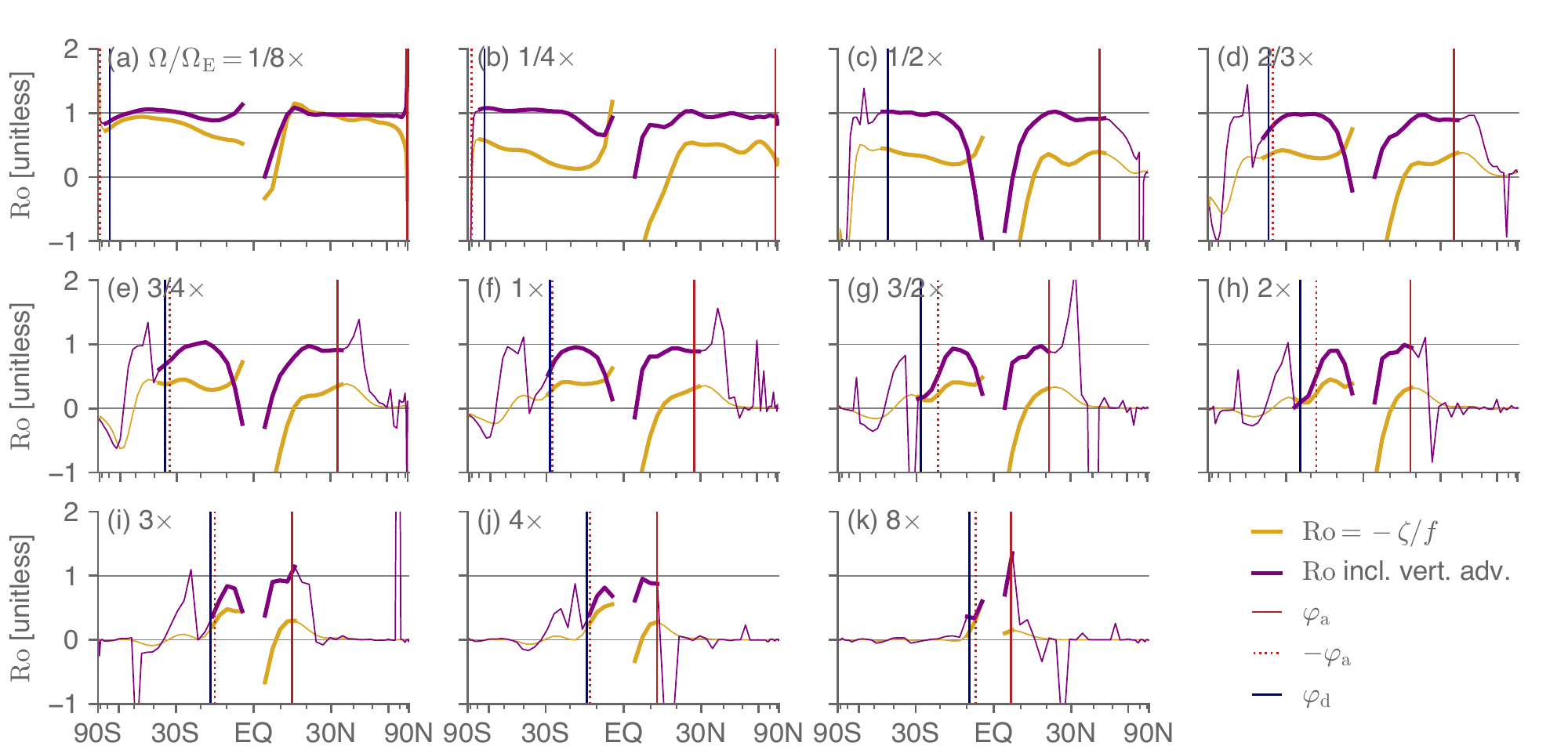}\\
  \caption{Rossby number in the \citet{singh_limits_2019} simulations at the 300~hPa level, computed either conventionally using \eqref{eq:ross-num} or the generalized form \eqref{eq:ross-num-gen} that accounts for the tilting of streamlines.  Overlaid are the cell edges \(\ascentlat\) and \(\latd\), with \(-\ascentlat\) also shown to ease comparison of the relative poleward extents of \(\ascentlat\) and \(\latd\).  Rossby number values outside of the Hadley circulation are shown as thinner curves, since they are less relevant, and they are masked out within 2\degr{} of the equator where division by the Coriolis parameter makes them less meaningful.}
  \label{fig:ro-s19-sims}
\end{figure*}

Finally, to justify our assumption in the scaling for \(\latd\) that the upper-tropospheric \(\Ro\) within the Hadley cell can be treated as fixed, Fig.~\ref{fig:ro-s19-sims} shows the meridional profile of \(\Ro\) at 300~hPa in each simulation of \citet{singh_limits_2019}, computed both conventionally as \eqref{eq:ross-num} and, following \citet{singh_limits_2019}, in a generalized form that incorporates vertical advection of angular momentum:
\begin{equation}
  \label{eq:ross-num-gen}
  \Ro_\mr{gen}\equiv\dfrac{1}{f}\left[-\zeta+\dfrac{\omega}{v}\pd{u}{p}\right],
\end{equation}
where \(v\) is meridional velocity, \(\omega\) is vertical velocity in pressure coordinates, \(p\) is pressure, and all quantities are zonal averages.  This accounts for the considerable tilting of streamlines, which causes the conventional \(\Ro\) to deviate from unity even if streamlines and angular momentum contours are everywhere parallel.  Though it is the conventional \(\Ro\) that appears in \(u_\Ro\) and thus ultimately our expressions for \(\latd\), for the simulations we argue that \eqref{eq:ross-num-gen} is more instructive: for the two-layer model of baroclinic instability utilized, the bulk upper-tropospheric behavior is more relevant than that at any chosen pressure level.  And as streamlines begin tilting toward the surface in the descending branch, the conventional \(\Ro\) at any given level decreases, while the bulk zonal velocities of the upper branch still roughly correspond to the planetary angular momentum values from where the streamlines exited the boundary layer in the ascending branch.  In other words, along streamlines angular momentum is nearly conserved (see Fig.~7 of \citet{singh_limits_2019}), which the meridional profile of \(\Ro\) at a fixed pressure level cannot capture.

The generalized Rossby number is close to unity over a large fraction of the cross-equatorial Hadley cell extent in all cases.  (Both forms are masked out over 2\degr{}S-2\degr{}N where division by the Coriolis parameter makes them less physically meaningful.)  The difference made by the vertical advection term is particularly large in the ascending branches.  For either version, we subjectively identify two regimes over the descending branch.  Slowly rotating cases have \(\Ro_\mr{gen}\) relatively uniform or even increasing slightly from the equator to the winter descending edge.  More rapidly rotating cases have \(\Ro_\mr{gen}\) decreasing poleward, approaching zero in the vicinity of the winter descending edge, but there is considerable scatter in the value of \(\Ro_\mr{gen}\) at the edge.  Despite this variation in the Rossby number across the simulations, it is evidently small enough that taking the bulk \(\Ro_\mr{gen}\) value as fixed in our scalings does not introduce major error.

\section{Conclusions}
\label{sec:conc}

\subsection{Summary}

We have introduced a new theory for the latitude of the poleward, descending edge of either Hadley cell (\(\latd\)) by combining the uniform-Rossby-number model of KL12---which uses a diagnosis of the ascending edge latitude (\(\ascentlat\)) to generalize the baroclinic instability-based model of H00 from annual-mean conditions to the annual cycle---with our recently presented theory for the ascending edge being set by the meridional extent of supercritical forcing \citep{hill_solsticial_2021}.  The theory predicts that \(\latd\) is displaced poleward when \(\Ro\) decreases or as \(\ascentlat\) moves poleward, and \(\ascentlat\) varies with the thermal Rossby number to the one-third power.  But in the small-angle limit reasonable for Earth, the dependence on \(\ascentlat\) drops out and the scaling for \(\latd\) predicts a one-fourth power dependence on the planetary Burger number, or equivalently on the thermal Rossby number provided the \(\deltav/\deltah\Ro\) remain fixed.

In an Earth-like, seasonally forced idealized aquaplanet simulation with a relatively shallow, 10-m mixed layer ocean depth, \(\ascentlat\) migrates rapidly to \(\sim\)25\degr{} into either summer hemisphere, and this seasonal cycle is well captured by the supercriticality-based scaling.  The summer cell is too weak for the summer \(\latd\) to be meaningful, but the winter \(\latd\) varies by only \(\lesssim5^\circ\) latitude about its mean position in either hemisphere.  Our combined theory predicting \(\ascentlat\) and \(\latd\) captures this behavior with \(\Ro\) kept at unity as in the original H00 model, but requires in place of \(\Ro\) variations that the \(\latd\) prediction be lagged by one month from that of \(\ascentlat\).

In simulations across a wide range of planetary rotation rates in three idealized GCMs, both \(\latd\) and \(\ascentlat\) adhere to the respective power-law exponents predicted by our theories in the relevant small thermal Rossby number regime.  This, combined with a smaller proportionality constant for \(\latd\) compared to \(\ascentlat\), helps explain why at very slow rotation rates the solsticial Hadley cell ascends essentially at the summer pole but considerably equatorward of the winter pole, \(\sim\)70\degr{}, rather than being roughly symmetric in extent about the equator as for more rapidly rotating cases.



\subsection{Discussion}

How might a predictive theory for \(\Ro\) be constructed?  \citet{hoskins_detailed_2020} offer an intriguing perspective relating to the frequency of deep convection in the ascending branch.  They argue that only when convection is sufficiently deep will there be upper-tropospheric meridional outflow that travels nearly inviscidly (\ie/ with \({\Ro\approx1}\)) toward either pole; at times and longitudes where deep convection is absent, they argue \({\Ro\approx0}\).  Under those conditions, the time-mean, zonal-mean \(\Ro\) field becomes a function of the spatial and temporal occurrence of deep convection in the ascending branch.  This contrasts with the conventional, extratropically focused approach to \(\Ro\), wherein it is controlled by stresses from subtropical and mid-latitude eddies propagating into the deep tropics and breaking \citep{walker_eddy_2006,schneider_general_2006}.

\citet{vallis_response_2015} speculate that Rossby waves are generated at the latitude of baroclinic instability onset, that these Rossby waves then propagate equatorward and break, and that the Hadley cell terminates at this wave-breaking latitude rather than the instability onset latitude.  This equatorward displacement may relate to our need for the \(c_\mr{d}<1\) parameter value to fit the \(\latd\) annual cycle in the seasonally forced aquaplanet simulation.  And the overall dynamics may help explain the 1-month lag for \(\latd\) relative to \(\ascentlat\) required to capture the precise month-to-month behavior of \(\latd\): it takes a finite amount of time for the zonal winds in the descending branch to adjust to a change in \(\ascentlat\) in the opposite hemisphere, and so too for the Rossby wave development, propagation, and breaking.

At the same time, across rotation rates the best fit \(c_\mr{d}\) parameters are greater than unity in some cases (Table~\ref{table:power-fits}), which is hard to square with this Rossby wave-based mechanism of \citet{vallis_response_2015}.  Moreover, the physical credibility of the two-layer model's critical shear criterion for baroclinic instability has been rightly questioned; a series of studies utilize a more comprehensive treatment of baroclinic instability to argue that \(\latd\) occurs where the vertical extent of baroclinic eddies spans a sufficient fraction of the troposphere \citep{korty_extent_2008,levine_response_2011,levine_baroclinic_2015}.  The same studies also incorporates the influence of moisture on the effective static stability, \ie/ on \(\deltav\) \citep{levine_response_2011,levine_baroclinic_2015}.

Though we have relied on \(\Ro\) being uniform over the upper branch of each Hadley cell (\cf/ KL12), the baroclinic instability criterion is computed latitude-by-latitude, and as such strictly speaking the behavior of \(\Ro\) equatorward of the instability onset latitude is irrelevant.  This also contrasts with the equal-area model appropriate for axisymmetric atmospheres, which depends on a measure of the potential temperature field integrated meridionally over the expanse of the cell.


Under annual-mean forcing in two dry and one moist idealized GCM, \citet{mitchell_constraints_2021} find that \(\latd\) scales as \(\Omega^{-1/3}\) in all three models.  This could be squared with our \(\Omega^{-1/4}\) scaling for \(\ascentlat=0\) if \(\Ro\) scales as \(\Omega^{-2/3}\).  By eye from their Fig.~8 and 10, \(\Ro\) does indeed follow an exponent close to this in two of the models---the same dry GCM we use and the moist GCM used by \citet{faulk_effects_2017}.  But a simpler dry dynamical core \citep{held_proposal_1994} shows no clear dependence of \(\Ro\) on \(\Omega\).   \citet{mitchell_constraints_2021} also put forward an ``omega governor'' mechanism which operates in the case that static stability and the effective heating (diabatic plus eddy heat convergence) averaged over the descending branch do not change.  Under those conditions, the poleward extent and mass overturning rate of the Hadley cell must vary in tandem: the cell weakens if it narrows, and it widens if it strengthens.  Prior to any adjustment by \(\latd\), if \(\ascentlat\) moves poleward then the cell widens, which under the omega governor would act to strengthen the overturning.  One can imagine that strengthening causing \(\Ro\) to increase, insofar as parcels then traverse the upper branch more rapidly and hence are less exposed to eddy stresses.  The increase in \(\Ro\) would, all else equal, act to move \(\latd\) equatorward, countering the direct influence of \(\ascentlat\) moving poleward.

Our theory could be further tested in numerous ways: against reanalysis data for the climatological annual cycle of the Hadley cells, against reanalysis data for interannual variability and trends, against comprehensive climate model simulations of global warming (\cf/ KL12), and against simulations of other terrestrial planetary atmospheres.

\acknowledgments
We are grateful to Sean Faulk and Martin Singh for sharing the data from their simulations and for many valuable discussions.  S.A.H. acknowledges financial support from NSF Award 1624740 and from the Monsoon Mission, Earth System Science Organization, Ministry of Earth Sciences, Government of India.  J.L.M. acknowledges funding from the Climate and Large-scale Dynamics program of the NSF, Award 1912673.

\bibliographystyle{ametsoc2014}
\bibliography{./references}

\end{document}